\documentclass[aps,prb,reprint,twocolumn,showpacs,floatfix,superscriptaddress,nofootinbib]{revtex4}
\usepackage{graphicx}
\usepackage{amssymb}
\usepackage{amsmath}
\usepackage{lmodern}
\usepackage{color}
\usepackage{hyperref}
\usepackage{epstopdf}
\begin{document}

\title{Spin-anisotropic magnetic impurity in a Fermi gas: poor man's scaling equation  integration}

\author{Eugene Kogan}
\email{Eugene.Kogan@biu.ac.il}
\affiliation{Jack and Pearl Resnick Institute, Department of Physics, Bar-Ilan University, Ramat-Gan 52900, Israel}
\affiliation{Computational Quantum Matter Research Team,
RIKEN Center for Emergent Matter Science (CEMS), Wako, Saitama 351-0198, Japan}

\author{Kazuto Noda}
%\email{kazuto.noda@riken.jp}
\affiliation{Computational Quantum Matter Research Team,
RIKEN Center for Emergent Matter Science (CEMS), Wako, Saitama 351-0198, Japan}

\author{Seiji Yunoki}
%\email{yunoki@riken.jp}
\affiliation{Computational Quantum Matter Research Team,
RIKEN Center for Emergent Matter Science (CEMS), Wako, Saitama 351-0198, Japan}
\affiliation{Computational Condensed Matter Physics Laboratory, RIKEN, Wako, Saitama 351-0198, Japan }
\affiliation{Interdisciplinary Theoretical Science (iTHES) Research Group, RIKEN, Wako, Saitama 351-0198, Japan}
\affiliation{Computational Materials Science Research Team,
RIKEN Advanced Institute for Computational Science (AICS), Kobe, Hyogo 650-0047, Japan}
\date{\today}

\begin{abstract}
We consider a single magnetic impurity described by the spin--anisotropic s-d(f) exchange (Kondo) model
and formulate   scaling equation for the spin-anisotropic model  when  the density of states (DOS)  of electrons is a power law function of energy (measured relative to the Fermi energy).
 We solve this equation
containing terms up to the second order in coupling constants in terms of elliptic functions.
From the obtained solution we find
the phases corresponding   to the infinite isotropic antiferromagnetic Heisenberg exchange,  to the impurity spin  decoupled from the electron environment (only for the pseudogap DOS), and   to the infinite Ising   exchange (only for the diverging DOS). We analyze
 the critical surfaces, corresponding to the finite  isotropic antiferromagnetic Heisenberg exchange for the pseudogap DOS.

\end{abstract}

\pacs{75.50.Mm, 72.15.Qm, 03.75.Mn}

\maketitle

\section{Introduction}

Kondo model, describing a magnetic moment
embedded in a system of non-interacting fermions continues to attract attention of both theorists and experimentalists for more than half a century.
In spite of  the model seeming simplicity, it is well known that the model is as far from being simple, as one can get, especially when additional complicated factors, like spin-anisotropy or non-flat behaviour of the electrons density of states (DOS) are to be taken into account.
 A very insightful approach to  this model is based on scaling equation pioneered by Anderson \cite{anderson,hewson}.
As far as spin-anisotropic model is concerned, to the best of our knowledge only the case of the $XXZ$ model (see below) with the flat DOS was analysed
in a detailed way \cite{hewson}.

A pseudogap Kondo model,  with a
power-law  DOS $\rho(\epsilon)\sim|\epsilon|^r$, which
has recently attracted a lot of attention \cite{sengupta,wehling,vojta,uchoa}
(for review, see Ref. \onlinecite{fritz}). In particular, graphene,
where the Kondo effect was observed recently \cite{chen}, is considered
as a typical realization of this  model.
More generally, one is interested in  Kondo problem for spin coupled to electrons with the pseudogap
or diverging  DOS \cite{vojta3,zhuravlev,kanao,cazalilla,mitchell2,mitchell,shirakawa}.

A model of a Kondo-like impurity interacting spin-isotropically with a band of fermions for which the DOS is zero or small near the Fermi energy was considered in Ref. \onlinecite{fradkin}.
There renormalization-group arguments were used to demonstrate that this model has a nontrivial zero-temperature phase
transition at a finite coupling constant, in contrast to the zero-coupling-constant transition of the
Kondo model with  constant DOS.
 Numerical and perturbative
renormalization study on Kondo and Anderson
models have provided a comprehensive understanding of
phase diagrams and thermodynamic properties \cite{fritz, chen2,buxton,bulla,bulla2,glossop,fritz2}.
However, a spin-anisotropic pseudogap Kondo model
to our best knowledge was never  considered.

The rest of the paper is constructed as follows.
We formulate in Section \ref{poor}   poor man's scaling equation for the spin-anisotropic model
with the power law DOS. In
Section \ref{XXZm}    solutions of scaling equations satisfying $J_x=J_y$  condition, referred to as  $XXZ$ model, are presented
as a preparation for the  Section \ref{xyzm}, where  we integrate the general scaling equation for the spin-anisotropic model.
Some important mathematical details and geometric  interpretation  of the  solution  are relegated to the Appendix.

\section{Poor man's scaling for the spin-anisotropic model}
\label{poor}

\subsection{Hamiltonian and scaling equation}

Poor man's scaling is the renormalization idea applied to the model of a single magnetic impurity in the Fermi sea of itinerant electrons.
The  Hamiltonian of the model can be written as
\begin{eqnarray}
\label{hamilto}
H=H_0+V=\sum_{{\bf k}\alpha}\epsilon_{\bf k}c_{{\bf k}\alpha}^{\dagger}c_{{\bf k}\alpha}+V,
\end{eqnarray}
where $c^{\dagger}_{{\bf k}\alpha}$ and $c_{{\bf k}\alpha}$ are  electron creation and annihilation operators,  $\epsilon_{\bf k}$ is the energy of itinerant electron with  wave vector ${\bf k}$ and spin $\alpha$, and the operator $V$ describes interaction between the electrons and the  impurity.

To formulate the renormalization procedure we need three objects: the Hamiltonian of the system $H$, the Hilbert space ${\cal H}$
(which is the product of the band of  itinerant electrons of width $D$ and of the Hilbert space where the impurity lives), and the $T$ matrix,
given by the series
\begin{eqnarray}
\label{pertu}
T=V+VG_0V+VG_0VG_0V+\dots,
\end{eqnarray}
where $G_0=(E+i0-H_0)^{-1}$.

Suppose we are interested only in the matrix elements of the $T$ matrix between the electron states at a distance from  the Fermi energy much less than the band width. Can we ignore the band edges? The answer is ``No'', because of virtual transitions of the electrons to the band edges, and the virtual transitions is all what quantum mechanics is about. Can we take into account all virtual transitions using  perturbative expansion (\ref{pertu})? The answer is again ``No'', because the series diverges due to transitions to the stats close to the Fermi energy.

The brilliant idea of Anderson was that we can take into account virtual transition to the band edges perturbatively,
that is by taking into account only a few first terms in Eq. (\ref{pertu}),
thus
reducing the band width $D$ of the itinerant electrons and calculating
the renormalization of the Hamiltonian  due to the elimination of the above mentioned virtual transitions.
Thus we reduce the Hilbert space ${\cal H}$ and renormalize the Hamiltonian $H$ accordingly, to keep the $T$
matrix constant.
And we can repeat this procedure again and again.

Now let us consider Kondo model. We find it appropriate to write down the   exchange part of the Hamiltonian
in explicitly rotation invariant form
\begin{eqnarray}
\label{hamiltonian}
V=H_{\text{ex}}=\sum_{ij}J_{ij}S^is^j(0),
\end{eqnarray}
where
$\vec{S}$ is the (siting at ${\bf r}=0$) impurity  spin operator (spin  is   one half),
$\vec{s}(0)=\frac{1}{2}\sum_{{\bf k}{\bf k}'\alpha\beta}c_{{\bf k}'\alpha}^{\dagger}\vec{\sigma}_{\alpha\beta}c_{{\bf k}\beta}$,
($\sigma^x,\sigma^y,\sigma^z$ are  Pauli matrices) is the itinerant electrons spin operator, and $J_{ij}$ are the anisotropic exchange coupling constants.

Let us consider first the traditional case of  flat DOS.
Poor Man's scaling consists in reducing the band width $D$ of the itinerant electrons and calculating perturbatively
the renormalized interactions due to the elimination of the virtual excitations to the band edges.
For the isotropic model
\begin{eqnarray}
\label{hamiltonianiso}
H_{\text{ex}}=J\vec{S}\cdot\vec{s}(0)
\end{eqnarray}
scaling   equation in the lowest order is  \cite{anderson,hewson}
\begin{eqnarray}
%\label{iso}
\frac{d J}{d\ln\Lambda}=-2\rho J^2,
\end{eqnarray}
where $\rho$ is DOS.

Consider   generalization of Hamiltonian (\ref{hamiltonian}):
\begin{eqnarray}
\label{hamiltoniansu}
H_{\text{ex}}=\sum_{ij=1}^{N^2-1}J_{ij}S^iT^j,
\end{eqnarray}
where  $S^i$ and $T^j$ are traceless generators of the group $SU(N)$.

Quadratic term in the scaling equation appears due to elimination of virtual transition of electron to the band edges in the lowest order of perturbation theory. Hence,
in the process of renormalization of the  anisotropic Hamiltonian there appears the factor
\begin{eqnarray}
\label{tensorsu}
&&J_{ij}S^iT^jJ_{kl}S^kT^l\\
&&=\frac{1}{4}J_{ij}J_{kl}\left(\{S^i, S^k\}\{T^j,T^l\}+[S^i, S^k][T^j,T^l]\right)\nonumber
\end{eqnarray}
(in  Eqs. (\ref{tensorsu})- (\ref{scalinga000}) we accept the summation convention: in all equations  summation  from 1 to $N^2-1$ with respect to every repeated index is implied).

Taking into account that
\begin{eqnarray}
\label{ksu}
\{S_i,S_k\}&=&\frac{\delta_{ik}}{N}+d_{ikn}S_n \nonumber\\
\left[S_i,S_k\right]&=&if_{ikn}S_n,
\end{eqnarray}
where $N\times N$  unit matrix is suppressed, the $d$-coefficients are symmetric in all indices,
and $f$ are structure constants, we obtain
\begin{eqnarray}
\label{tensor1}
&&4J_{ij}S^iT^jJ_{kl}S^kT^l=\frac{1}{N^2}J_{ij}J_{ij}\nonumber\\
&&+\frac{1}{N}J_{ij}J_{jk}d_{ikm}\left(S^m+T^m\right)\\
&&+J_{ij}J_{kl}\left(d_{ikm}d_{jln}-f_{ikm}f_{jln}\right)S^mT^n.\nonumber
\end{eqnarray}

Thus for  the $SU(2)$ Kondo model  one obtains \cite{cox,irkhin}
\begin{eqnarray}
\label{scalinga000}
\frac{d J_{mn}}{d\ln\Lambda}=-\rho J_{ij}J_{kl}\epsilon_{ikm}\epsilon_{jln},
\end{eqnarray}
where $\epsilon$ is Levi-Civita symbol.

The microscopic tensor $J_{ik}^{(m)}$   can always be reduced  to principal axes by rotation of the coordinate system, and it keeps it's
diagonal form  in the process of renormalization, as we see from Eq. (\ref{scalinga000}). So we can write down the exchange Hamiltonian as
\begin{eqnarray}
\label{h}
H_{\text{ex}}=\sum_iJ_{i}S^is^i(0),
\end{eqnarray}
and the scaling equation as
\begin{eqnarray}
\label{ggg}
\frac{d J_x}{d\ln\Lambda}&=&-2\rho J_yJ_z \nonumber\\
\frac{d J_y}{d\ln\Lambda}&=&-2\rho J_xJ_z \\
\frac{d J_z}{d\ln\Lambda}&=&-2\rho J_xJ_y,\nonumber
\end{eqnarray}
and sweep under the carpet the question of stability of the diagonal solution by adding to the poor man's scaling a possible rotation of coordinate system at each step. \footnote{Actually, the Hamiltonian (\ref{hamiltonian}) (or (\ref{h})) is more meaningful for the theory of two level systems, $S^i$ and $s^i$ being pseudospin operators \cite{cox}, than for the orthodoxal Kondo model \cite{nozieres}.}

Now let us return to the case when the electron dispersion law determines the power law dependence of the DOS upon the energy
\begin{eqnarray}
\label{e}
\rho(\epsilon)=C|\epsilon|^r,\;\;\;\text{if}\;\;|\epsilon|<D,
\end{eqnarray}
where $r$ can be either positive or negative ($r>-1$) \cite{mitchell}.
For the DOS we consider, in distinction from the standard renormalization procedure \cite{hewson},  one has additionally to rescale the unit of length \cite{fradkin}.
Thus for the isotropic model (\ref{hamiltonianiso})
scalar scaling   equation is \footnote{It is known that physics of the problem  can  change at $r=1/2$ \cite{fradkin,fritz}, and thus the range of validity of scaling equations (\ref{scalinga00}) is an open problem.
However, we do not discuss here this problem and admit the limitation of our approach: the value of $r$ determines just the scale of the problem (see below).}
\begin{eqnarray}
\label{iso}
\frac{d J}{d\ln\Lambda}=rJ-2GJ^2,
\end{eqnarray}
where  $G=CD^r$, and $\Lambda=D'/D$; $D'$ is the actual width of the itinerant electrons band after the exclusion of the virtual excitations to the
edges, was obtained and studied previously \cite{fradkin}.

Combining spin-anisotropy and power law DOS we should add linear terms to the RHS of Eq. (\ref{ggg}) (with zero $N$) to obtain
\begin{eqnarray}
\label{scalinga00}
\frac{d J_x}{d\ln\Lambda}&=&rJ_x-2GJ_yJ_z \nonumber\\
\frac{d J_y}{d\ln\Lambda}&=&rJ_y-2GJ_xJ_z \\
\frac{d J_z}{d\ln\Lambda}&=&rJ_z-2GJ_xJ_y.\nonumber
\end{eqnarray}

\subsection{Symmetries and fixed points of the scaling equation}
\label{sis}

Equation (\ref{scalinga000}) is symmetric with respect to all space rotations (the group $K$ \cite{landau3}), from which only the symmetry with respect to permutation of the indices $x,y,z$ is left for  Eq. (\ref{scalinga00}). Additionally, both   equations are  symmetric with respect to space inversion accompanied by the inversion of the direction of flow and change of  sign of $r$.
So further on  we consider explicitly only the case $r>0$ (and  present the results for negative $r$ in some cases).

We introduce $\lambda=\Lambda^r$ and, unless $G$ appear explicitly in the equation,  measure $J$  in units of $r/2G$.
So   Eq. (\ref{scalinga00}) becomes
\begin{eqnarray}
\label{scalinga0b}
\lambda\frac{dJ_x}{d\lambda}&=&J_x-J_yJ_z\nonumber\\
\lambda\frac{dJ_y}{d\lambda}&=&J_y-J_xJ_z\\
\lambda\frac{dJ_z}{d\lambda}&=&J_z-J_xJ_y.\nonumber
\end{eqnarray}
When we  look for the flow lines of Eq. (\ref{scalinga0b}),   the parameter $\lambda\in(0,+\infty)$ (and decreases along
a flow line). When we  consider the  physical problem,   the parameter $\lambda\in (0,+1]$, and Eq. (\ref{scalinga0b}) becomes the initial (final) value problem with
\begin{eqnarray}
\label{subs}
J_n(1)=2GJ_n^{(m)}/r.
\end{eqnarray}

Equation (\ref{scalinga0b}) has a trivial fixed point
\begin{eqnarray}
\label{zero}
J_x^*=J_y^*=J_z^*=0,
\end{eqnarray}
 corresponding to the impurity spin  decoupled from the electron environment, and  four non-trivial ones
\begin{eqnarray}
\label{odd}
|J_x^*|=|J_y^*|=|J_z^*|=1;\;\;\; \;\;\;J_xJ_yJ_z=1,
\end{eqnarray}
corresponding to finite isotropic antiferromagnetic Heisenberg exchange.

Apart from the finite fixed points given by Eqs. (\ref{zero}) and (\ref{odd}),  Eq. (\ref{scalinga0b}) has infinite fixed points (more precisely, rays starting at the origin and going to the infinity, which serve as attractors for the flow lines). However, it will be more convenient to discuss these attractors later (in the Subsection \ref{tt}).

 It is obvious that the  trivial fixed point  is stable. (We remind that we consider here only the case of positive $r$.)
To analyze stability of the non-trivial fixed points,  we write $J_n=J_n^*+\delta J_n$ and linearize Eq. ({\ref{scalinga0b})   with respect to deviations from the fixed point $\delta J_n$. Thus we obtain
\begin{eqnarray}
\label{scaling2}
\frac{d \delta J_n}{d\lambda}=\sum_mT_{nm}\delta J_m,
\end{eqnarray}
where the eigenvalues of  matrix $T$ for any fixed point are $-1$ and doubly degenerate $2$. Hence all the non-trivial fixed points are semi-stable and, hence, are critical points.

Taking the second step we introduce
$\widetilde{J}_n=J_n/\lambda$,
and  Eq. (\ref{scalinga00}) takes the form
\begin{eqnarray}
\label{scalinga0}
\frac{d \widetilde{J}_x}{d\lambda}&=&-\widetilde{J}_y\widetilde{J}_z\nonumber\\
\frac{d \widetilde{J}_y}{d\lambda}&=&-\widetilde{J}_x\widetilde{J}_z\\
\frac{d \widetilde{J}_z}{d\lambda}&=&-\widetilde{J}_x\widetilde{J}_y.\nonumber
\end{eqnarray}

\section{Integration of scaling equations for the $XXZ$ model}
\label{XXZm}

\subsection{What we can learn from isotropic model}

We start our analysis from the simple case of
isotropic model ($J_x=J_y=J_z=J$), though it was analyzed  before. However we prefer to reproduce the analysis, because this way we understand the pattern, which
will repeat itself throughout the paper.
Eq. (\ref{scalinga0}) in this case can be solved immediately
\begin{eqnarray}
\label{elem}
\widetilde{J}=\frac{1}{\lambda+\psi}\Longrightarrow J=\frac{\lambda}{\lambda+\psi}.
\end{eqnarray}

From the point of view of a mathematician, $\psi$ is just the constant of integration.
From the point of view of a physicist
$\psi$ for the particular problem has the particular value, connected with microscopic parameters by Eq. (\ref{subs}).
For $\psi>0$, when  $\lambda$ decreases to zero,
 $J(\lambda)$ converges to the trivial fixed point, which means that the spin is decoupled from the environment (when energy goes to zero). The value of $\psi=0$ means a critical point, that is energy independent interaction of the spin with the environment. And finally,
if $-1<\psi<0$, then $J(\lambda)$ has a relevant pole  at some finite
value of $\lambda$, corresponding to finite value of $D'$. (This pole  is similar to Landau pole \cite{landau}.)

Formally, this pole just means that the perturbation theory (and scaling equation  (\ref{scalinga00}) is a clever but still perturbation theory)
breaks down. On the other hand, as it is known since long ago, the value of $D'$ mentioned above
provides an estimate of Kondo temperature $T_K$
\begin{eqnarray}
\label{Kondo}
T_K\sim D'=D(-\psi)^{1/r}.
\end{eqnarray}
So the parameter $\psi$  has a clear physical meaning.

Substituting Eq. (\ref{elem})
 into Eq. (\ref{subs})  one obtains \cite{mitchell2}
\begin{eqnarray}
\label{Kondo2}
T_K\sim D\left(1-\frac{r}{2GJ^{(m)}}\right)^{1/r},
\end{eqnarray}
provided
\begin{eqnarray}
0<\frac{r}{2GJ^{(m)}}<1.
\end{eqnarray}
Note that if we take the limit $r\to 0$ in Eq. (\ref{Kondo2}), we obtain
\begin{eqnarray}
T_K\sim De^{-1/2GJ^{(m)}}.
\end{eqnarray}

\subsection{The $XXZ$ model}
\label{ing}

Now let us go to the $XXZ$ model ($J_x=J_y$).   Eq. (\ref{scalinga0}) in this case takes the form
\begin{eqnarray}
\label{scalinga98}
\frac{d \widetilde{J}_x}{d\lambda}&=& -\widetilde{J}_z \widetilde{J}_x \nonumber \\
\frac{d  \widetilde{J}_z}{d\lambda}&=&-\widetilde{J}_x^2.
\end{eqnarray}
We immediately obtain the first integral of  Eq.  (\ref{scalinga98})
\begin{eqnarray}
\label{scalingac2}
\widetilde{J}_x^2-\widetilde{J}_z^2=\pm A^2.
\end{eqnarray}
Substituting  into Eq. (\ref{scalinga98}) and integrating
we get
\begin{eqnarray}
\label{d1b}
J_x&=&\pm A\lambda\cdot\mathrm{csc(h)}(A\lambda+\psi) \nonumber\\
J_z&=&A\lambda\cdot\mathrm{cot(h)}(A\lambda+\psi).
\end{eqnarray}
 In Eq. (\ref{d1b}) $\mathrm{cos(h)}$ stands for either trigonometric or hyperbolic cosine, and similar for $\mathrm{cot(h)}$.

 Note that presence of two integration constants ($\psi$ and $A$) in the solution (\ref{d1b}) reflects two symmetries of Eq. (\ref{scalinga0}): with respect to transformation $\lambda\to\lambda+\psi$ and with respect to transformation $\lambda\to A\lambda,\widetilde{J}\to \widetilde{J}/A$.
  For trigonometric functions $\psi\in(-\pi/2,\pi/2]$, for hyperbolic functions
 $\psi\in(-\infty,+\infty)$.

Together with  fixed points, separatrices form the skeleton of a flow diagram. In our case non-trivial separatrices are
 described by putting $\psi=0$  in Eq. (\ref{d1b}). Thus we get
four  separatrices (ending at the critical points), described  by the equations
\begin{eqnarray}
\label{s1a}
\frac{\widetilde{J}_z}{|\widetilde{J}_x|}&=&\mathrm{cos(h)}\left(\sqrt{|\widetilde{J}_x^2-\widetilde{J}_z^2|}\right)
\end{eqnarray}
(in the  case of $\cos$ the solution of Eq. (\ref{s1a}) should also satisfy $\sqrt{\widetilde{J}_x^2-\widetilde{J}_z^2}<\pi$).
The asymptotic of the solution of Eq. (\ref{s1a}) is
\begin{eqnarray}
\widetilde{J}_x&=&\pm\left(\widetilde{J}_z-\frac{\pi^2}{2\widetilde{J}_z}\right),\;\;\;\widetilde{J}_z\ll -1  \nonumber\\
\widetilde{J}_x &=& \pm \widetilde{J}_ze^{-\widetilde{J}_z},\;\;\;  \widetilde{J}_z\gg 1.
\end{eqnarray}
The trivial separatrices  are $\widetilde{J}_x=0$  and
 $ \widetilde{J}_x=\pm \widetilde{J}_z$.

A flow diagram, as described by   Eq. (\ref{d1b}), is shown in Fig. \ref{XXZ}. Because of the symmetry of Eq. (\ref{scalinga0b}) it is enough to plot only the upper part of the phase plain $J_x\geq 0$. We observe the non-interacting phase, corresponding to the
trivial fixed point, the phase of infinite isotropic antiferromagnetic Heisenberg exchange,
(both phases corresponding to stable fixed points) We also observe
 the critical  line of finite isotropic antiferromagnetic Heisenberg exchange, ending at the critical point (semi-stable fixed point). This Figure shows an example of asymptotic symmetry \cite{pokrovskii}. After the renormalization the system becomes isotropic (or trivial) even it was anisotropic microscopically.

The same flow diagram as in Fig. \ref{XXZ}, is shown in Fig. \ref{large}, but this time the plot includes larger values of  $J_x,J_z$
(or smaller values of $r$). The main purpose of this Figure is to illustrate how our results are reduced to those obtained
for the constant DOS \cite{hewson} when $r\to 0$.
For large $J_x,J_z$ the linear terms in Eq. (\ref{scalinga98}) can be neglected, and  the flow diagram naturally looks like the one
from Ref. \onlinecite{hewson}, which  consists
of hyperbolas.
However, when in the process of evolution at least one of $J_x,J_z$ becomes of order  one, the principal deviations from the hyperbolas can be clearly seen.
Fig. \ref{large} also illustrates how the fixed line $J_x=0$ obtained for $r=0$  \cite{hewson} emerges when $r\to 0$.

Flow diagram for negative $r$, presented on Fig. \ref{negative}, is just the diagram from Fig. \ref{XXZ}, replotted taken into account the  symmetry
of the problem mentioned in the beginning of the Subsection \ref{sis}. Because change of sign of $r$ should be accompanied by the inversion of the direction of flow (and space inversion), after this change the previously stable fixed points  become unstable (and of no physical interest) and vice versa. Thus for negative $r$ we have two phases, corresponding respectively to  infinite isotropic antiferromagnetic Heisenberg exchange and  infinite   Ising exchange. We want to emphasise again, that physical fixed points for negative $r$ correspond to nonphysical fixed points for positive $r$.
Notice also that for negative $r$ the critical point is  ferromagnetic (and the critical line is totally different from that for positive $r$). Thus Fig. \ref{negative}  shows (depending upon the initial conditions) either the same asymptotic symmetry we had for positive $r$ or  dynamical
 generation of  anisotropy.
\begin{figure}[h]
\vskip -.5cm
\hskip -.5cm
\includegraphics[width= 1.05\columnwidth]{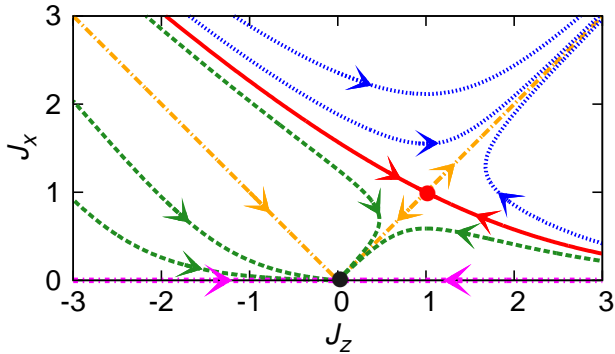}
\vskip -1cm
\caption{(color online) Flow diagram as described by  Eq. (\ref{d1b}). Trivial fixed point is shown by black circle, critical  point  by red circle,  isotropic model  by orange dot  dashed line, Ising model  by violet dot dot dashed line, critical line  by red solid line. The   non-interacting (infinite isotropic antiferromagnetic Heisenberg exchange) phase is shown by green dashed (blue dotted) lines. }
 \label{XXZ}
\end{figure}
\begin{figure}[h]
\vskip -1cm
\hskip -.5cm
\includegraphics[width= 1.05\columnwidth]{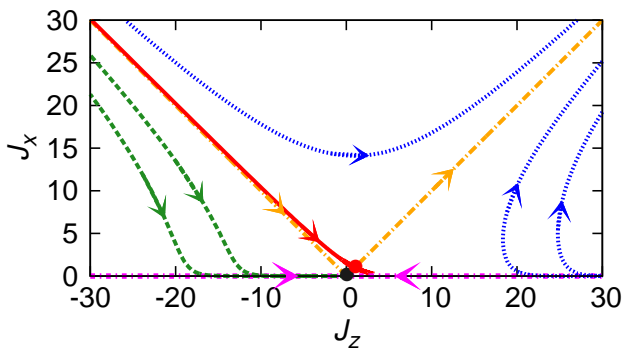}
\vskip -1cm
\caption{ (color online) Same as  Fig. \ref{XXZ}, but with a wider plot interval. }
\label{large}
\end{figure}

\begin{figure}[h]
\vskip -1cm
\hskip -.5cm
\includegraphics[width= 1.05\columnwidth]{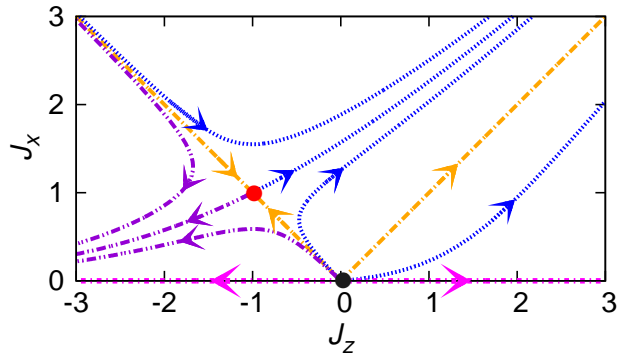}
\vskip -1cm
\caption{(color online) Flow diagram for negative $r$.
The  infinite  Ising exchange  phase is shown by  magenta dot short dashed lines.}
\label{negative}
\end{figure}

As far as Kondo effect is concerned,  we can repeat verbatim the two paragraphs following  Eq. (\ref{elem}), only   $\psi$ in the RHS of Eq. (\ref{Kondo}) should be divided by $A$. The values of $\psi$ and $A$ are found this time by substituting  Eq. (\ref{d1b}) into Eq.  (\ref{subs}).
\footnote{We recently learned that part of results of the Subsection \ref{ing} were independently  obtained in Ref. \onlinecite{ingersent}.}

\section{Integration of scaling equations in the general case}
\label{xyzm}

\subsection{Tale of two integrals}
\label{tt}

Let us generalize Eq. (\ref{scalinga0}) to
\begin{eqnarray}
\label{scalinga0g}
\frac{d \widetilde{J}_x}{d\lambda}&=&Q\widetilde{J}_y\widetilde{J}_z\nonumber\\
\frac{d \widetilde{J}_y}{d\lambda}&=&R\widetilde{J}_x\widetilde{J}_z\\
\frac{d \widetilde{J}_z}{d\lambda}&=&S\widetilde{J}_x\widetilde{J}_y,\nonumber
\end{eqnarray}
where $Q,R,S$ are some constants. Equation (\ref{scalinga0g}) includes the $XYZ$ model ($P=Q=R=-1$), and also other, much more important cases, like Euler top \cite{landau2}.

From Eq. (\ref{scalinga0g}) follows
\begin{eqnarray}
\label{scaling}
\frac{d}{d\lambda}\left(a\widetilde{J}_x^2+b\widetilde{J}_y^2+c\widetilde{J}_z^2 \right)=0,
\end{eqnarray}
where $a,b,c$ are arbitrary constants satisfying
\begin{eqnarray}
\label{pqr}
aQ+bR+cS=0.
\end{eqnarray}
Hence we immediately get two first integrals
\begin{eqnarray}
\label{pro2}
a_1\widetilde{J}_x^2+b_1\widetilde{J}_y^2+c_1\widetilde{J}_z^2&=&A_1\nonumber\\
a_2\widetilde{J}_x^2+b_2\widetilde{J}_y^2+c_2\widetilde{J}_z^2&=&A_2,
\end{eqnarray}
where  $(a_1,b_1,c_1)$ and $(a_2,b_2,c_2)$ should be linearly independent.
Equations (\ref{pro2}) seems to contain a lot of constants, but if we consider it (and Eq. (\ref{pqr}) as defining straight line in the space with the coordinates $(x,y,z)=(\widetilde{J_x}^2,\widetilde{J_y}^2,\widetilde{J_z}^2)$, we are motivated to present this equation in the canonical form
\begin{eqnarray}
\label{pro5}
\frac{x-x_0}{Q}=\frac{y-y_0}{R}=\frac{z-z_0}{S}.
\end{eqnarray}
Thus  in the space introduced above each flow lines  lies on the ray with the direction $(Q,R,S)$.

Returning temporarily to the Kondo problem we solve ($Q=R=S=-1$), we immediately understand by inspection of Eq. (\ref{pro5}) that the attractors of  Eq. (\ref{scalinga0b}) going to infinity
corresponds  to isotropic Hamiltonian.

\subsection{General solution}
\label{gs}

Two integrals being found, we are left with a single equation for a single variable ${\cal P}$,
which is naturally to chose according to the equation
\begin{eqnarray}
\label{pro55}
\frac{x-x_0}{Q}=\frac{y-y_0}{R}=\frac{z-z_0}{S}=\frac{\cal P}{QRS}.
\end{eqnarray}
Also, Eq. (\ref{pro2}) contains just two constants. To emphasize this fact we put on $x_0,y_0,z_0$ condition
\begin{eqnarray}
\label{constraint}
\frac{x_0}{Q}+\frac{y_0}{R}+\frac{z_0}{S}=0.
\end{eqnarray}
Substituting Eq. (\ref{pro55})  into Eq. (\ref{scalinga0g}) and taking into account Eq. (\ref{constraint}) we obtain \cite{shiba}
\begin{eqnarray}
\label{we}
\left[\frac{d{\cal P}(\lambda)}{d\lambda}\right]^2=4\left[{\cal P}(\lambda)\right]^3-g_2{\cal P}(\lambda)-g_3,
\end{eqnarray}
where
\begin{eqnarray}
\label{constraint2}
g_2&=&-4(RSx_0+QSy_0+QRz_0)\nonumber\\
g_3&=&-4Q^2R^2S^2x_0y_0z_0.
\end{eqnarray}
Hence ${\cal P}(\lambda)$ is Weierstrass elliptic function ${\cal P}(\lambda;\omega_1,\omega_2)$ \cite{abram}, and
 $\omega_1$, $\omega_2$ are connected with $g_2$, $g_3$ by equation
\begin{eqnarray}
\label{constraint3}
g_2&=&60\sum_{(m,n)\neq (0,0)}(m\omega_1+n\omega_2)^{-4} \nonumber \\
g_3&=&140\sum_{(m,n)\neq (0,0)}(m\omega_1+n\omega_2)^{-6}.
\end{eqnarray}

Thus we obtain the  solution of   Eq. (\ref{scalinga0b}) as
\begin{eqnarray}
\label{ammw}
J_x^2 &=&\lambda^2\left[{\cal P}(\lambda+\psi;\omega_1,\omega_2)/RS+x_0\right]\nonumber\\
J_y^2 &=&\lambda^2\left[{\cal P}(\lambda+\psi;\omega_1,\omega_2)/QS+y_0\right]\\
J_z^2 &=&\lambda^2\left[{\cal P}(\lambda+\psi;\omega_1,\omega_2)/QR+z_0\right].\nonumber
\end{eqnarray}
The solution represents a two-parameter ($\omega_1,\omega_2$) family of the flow lines,  with $x_0,y_0,z_0$ being connected with these two parameters by Eqs. (\ref{constraint}), (\ref{constraint2}), (\ref{constraint3}).

Alternative (and  more convenient) representation of the  solution of Eq. (\ref{scalinga0b})  through Jacobi elliptic functions is presented in the Appendix. The result for the problem we solve is
\begin{eqnarray}
\label{amm}
J_x &=&\pm A\lambda\cdot\mathrm{ns}(A\lambda+\psi,k)\nonumber\\
J_y &=&\pm A\lambda\cdot\mathrm{cs}(A\lambda+\psi,k)\\
J_z &=&\pm A\lambda\cdot\mathrm{ds}(A\lambda+\psi,k),\nonumber
\end{eqnarray}
where  the factors $\pm 1$ should satisfy condition that their product is equal to $+1$,  plus the
solutions which
can be obtained from Eq. (\ref{amm}) by interchanging $J_x,J_y,J_z$.

Equation (\ref{amm}) is the main result of the paper. It represents a two-parameter family of the flow lines. The  parameter  $\psi\in(-K(k),K(k)]$, where $K$ is the complete elliptic integral of the first kind,
and the parameter $k\in [0,1]$, and, as we show in the appendix, has the simple geometric meaning.

The results of the previous  Section are the particular case  of those obtained in this Section.
In fact, because
\begin{eqnarray}
\left\{\begin{array}{l}\mathrm{ns}(\phi,0)=\csc(\phi)\\
\mathrm{cs}(\phi,0)=\cot(\phi)\\
\mathrm{ds}(\phi,0)=\csc(\phi)\end{array}\right.\;\;\;
\left\{\begin{array}{l}\mathrm{ns}(\phi,1)=\coth(\phi)\\
\mathrm{cs}(\phi,1)=\mathrm{csch}(\phi)\\
\mathrm{ds}(\phi,1)=\mathrm{csch}(\phi)\end{array}\right.,
\end{eqnarray}
for $k=0$
Eq. (\ref{amm})  contains Eq. (\ref{d1b}) with trigonometric functions and $J_z$ and $J_y$ interchanged,
and for  $k=1$   it contains  the same equation   with hyperbolic functions and $J_z$ and $J_x$ interchanged.

The values of $J_x,J_y,J_z$ obtained from Eq. (\ref{amm}) for  $\lambda=0$ and $\psi \neq 0$ corresponds to  the trivial fixed point,  which describes the non-interacting phase. Further on we consider only  the case $r>0$.
For $\psi = 0$ the value $\lambda=0$ corresponds to one of  the critical points, and $\lambda\in(0,2K(k))$  corresponds to
the critical surface.

In Fig.\ref{surface}  we show the critical surface of one critical point.
 Due to the symmetry of equations mentioned in Sec. II,
three other critical surfaces can be obtained from presented on Fig. \ref{surface} by rotations by the angle $\pi$ about the Cartesian axes.
Fig. \ref{all} with all four critical surfaces gives
a complete picture of the phase diagram for the $XYZ$ model, with critical surfaces separating between the  infinite isotropic antiferromagnetic Heisenberg exchange phase and the phase, corresponding to the impurity spin  decoupled from the electron environment.

\begin{figure}[h]
\includegraphics[width= 1\columnwidth]{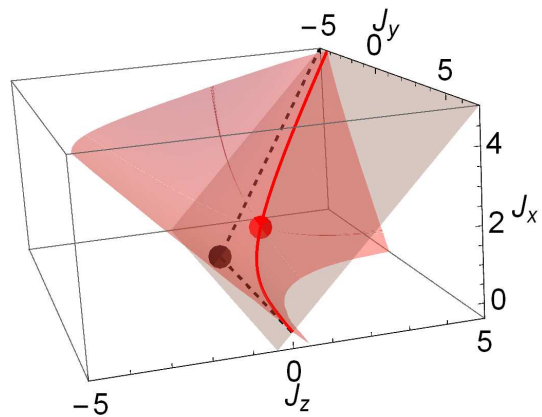}
\caption{  (color online) Critical surface of the critical  point $(J_x,J_y,J_z)=(1,1,1)$ is shown in red.
Solid (dotted) line is a critical line (asymptotes) on $J_x=J_z$ plane (painted in grey); these lines are the same as in Fig. \ref{XXZ}.}
\label{surface}
\end{figure}

\begin{figure}[h]
\includegraphics[width= .7\columnwidth]{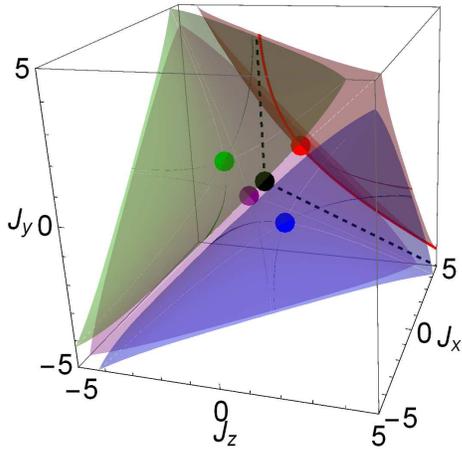}
\caption{  (color online)
All four critical surfaces.  Critical
points
are denoted by colored circles.}
\label{all}
\end{figure}

As far as Kondo effect is concerned,  we can repeat verbatim the  last paragraph of Section \ref{XXZm},
only this time Eq. (\ref{amm})  should be used instead of Eq. (\ref{d1b}).

\section{Conclusions}
\label{conclusions}

We considered a single magnetic impurity described by the spin--anisotropic s-d(f) exchange (Kondo) model.
We formulated the explicitly rotation invariant  scaling equation for the power law electron DOS
and solved this equation
 in terms of elliptic functions.

We found the  infinite isotropic antiferromagnetic Heisenberg exchange phase, the phase, corresponding to the impurity spin  decoupled from the electron environment (for the pseudogap DOS) and the infinite Ising exchange phase (for  the DOS diverging at the Fermi level). We studied in details   the critical surface
corresponding to the finite isotropic antiferromagnetic Heisenberg exchange (for the pseudogap DOS).

\begin{acknowledgments}

This work has been supported in part by RIKEN iTHES Project and Molecular Systems.
One of the authors (E.K.)  thanks RIKEN for the hospitality extended to him during
his stay.
The other author (K.N.) is supported by Grant-in-Aid for JSPS Fellows (Grant No. 16J07637).

The authors are grateful to N. Andrei, Y. Avishay, K. Ingersent, D. Khveshchenko, T. Kimura, Chi-Cheng Lee,  A. Mitchell, A. Nevidomskyy, Y. Ohyama, R. Sakano, Y. Teratani, and M. Vojta for valuable discussions,
and particularly to V. Yu. Irkhin for bringing to their attention Refs. \onlinecite{cox,irkhin} and to A. Oguri for bringing to their attention Refs. \onlinecite{shiba,yosida}.

\end{acknowledgments}

\begin{appendix}

\section{If you have seen one, you have seen them all}

We explained in Section \ref{poor} that if we know the quadratic terms in poor man's scaling equation for the Hamiltonian (\ref{hamiltonianiso}), we know  these terms in  poor man's scaling equation for the Hamiltonian (\ref{hamiltonian}). We claim that  this remains  true
for the higher order terms.
For example, from  scaling equation  for the isotropic Hamiltonian (\ref{hamiltonianiso}) \cite{hewson}
\begin{eqnarray}
\label{cube}
\frac{d J}{d\ln\Lambda}=-2\rho J^2+2\rho^2J^3,
\end{eqnarray}
 follows scaling equation  for  reduced anisotropic Hamiltonian
\begin{eqnarray}
\label{hamiltonianr}
H_{exc}=\sum_iJ_{i}S^is^i(0)
\end{eqnarray}
in the form
\begin{eqnarray}
\label{cu}
\frac{d J_x}{d\ln\Lambda}&=&-2\rho J_yJ_z+\rho^2(J_y^2+J_z^2)J_x \nonumber\\
\frac{d J_y}{d\ln\Lambda}&=&-2\rho J_xJ_z+\rho^2(J_x^2+J_z^2)J_y \nonumber\\
\frac{d J_z}{d\ln\Lambda}&=&-2\rho J_xJ_y+\rho^2(J_x^2+J_y^2)J_z.
\end{eqnarray}
This result can be obtained by considering expression
\begin{eqnarray}
\sum_{ikl}J_iS^i\sigma^i J_kS^k\sigma^kJ_lS^l\sigma^l
\end{eqnarray}
and applying twice Eq. (\ref{tensor1}).

\section{Solving Eq. (\ref{scalinga0g}) using Jacobi elliptic functions}

For convenience of the reader we present here basic facts concerning  Jacobi elliptic functions.
There are three major  functions:
$\mathrm{sn}(\lambda,k)$ solves the differential equation
\begin{eqnarray}
\left(\frac{du}{d\lambda}\right)^2=(1-u^2)(1-k^2u^2);
\end{eqnarray}
$\mathrm{cn}(\lambda,k)$ solves the differential equation
\begin{eqnarray}
\left(\frac{du}{d\lambda}\right)^2&=&(1-u^2)(1-k^2+k^2u^2);
\end{eqnarray}
$\mathrm{dn}(\lambda,k)$ solves the differential equation
\begin{eqnarray}
\left(\frac{du}{d\lambda}\right)^2=(1-u^2)(u^2-1+k^2).
\end{eqnarray}

Also, there are nine  minor  functions:
\begin{eqnarray}
\mathrm{pq}(\lambda,k)=\frac{\mathrm{pn}(\lambda,k)}{\mathrm{qn}(\lambda,k)},
\end{eqnarray}
where  $\mathrm{p}$ and $\mathrm{q}$ are any of the letters $\mathrm{n,s,c,d}$ ($\mathrm{nn}(\lambda)\equiv 1$),
named by  the first letter of the numerator followed by the first letter of the denominator.

The rules of differentiation of the elliptic functions are:
\begin{eqnarray}
\label{jac}
&&\left\{\begin{array}{l} \frac{d}{d\lambda}\mathrm{sn}=\mathrm{cn}\cdot\mathrm{dn} \\
\frac{d}{d\lambda}\mathrm{cn}=-\mathrm{sn}\cdot\mathrm{dn} \\
\frac{d}{d\lambda}\mathrm{dn}=-k^2\cdot\mathrm{sn}\cdot\mathrm{cn}
\end{array}\right.
\left\{\begin{array}{l} \frac{d}{d\lambda}\mathrm{nc}=\mathrm{sc}\cdot\mathrm{dc}\\
\frac{d}{d\lambda}\mathrm{sc}=\mathrm{nc}\cdot\mathrm{dc}\\
\frac{d}{d\lambda}\mathrm{dc}=(1-k^2)\cdot\mathrm{nc}\cdot\mathrm{sc}
\end{array}\right.\nonumber\\
&&\left\{\begin{array}{l}
\frac{d}{d\lambda}\mathrm{ns}=-\mathrm{cs}\cdot\mathrm{ds}   \\
\frac{d}{d\lambda}\mathrm{cs}=-\mathrm{ns}\cdot\mathrm{ds} \\
\frac{d}{d\lambda}\mathrm{ds}=-\mathrm{ns}\cdot\mathrm{cs}
\end{array}\right.\hskip .7cm
\left\{\begin{array}{l}
\frac{d}{d\lambda}\mathrm{nd}=k^2\cdot\mathrm{sd}\cdot\mathrm{cd} \\
 \frac{d}{d\lambda}\mathrm{sd}=\mathrm{cd}\cdot\mathrm{nd} \\
 \frac{d}{d\lambda}\mathrm{cd} =-(1-k^2)\cdot\mathrm{sd}\cdot\mathrm{nd}
\end{array}\right..\nonumber\\
\end{eqnarray}
 (the argument of all  functions is  $\lambda$, and the modulus is $k$).

These simple rules allow to integrate  Eq. (\ref{scalinga0g}) just by inspection. In fact,
there are four options for the signs of $P,Q,R$ in Eq. (\ref{scalinga0g}): three pluses, two pluses and one minus, one plus and two minuses, and three minuses. Equation (\ref{jac}) allows one
to solve Eq. (\ref{scalinga0g}) for each option; one should just read two previous paragraphs backward.
Thus Eq. (\ref{scalinga0}) is solved by the set of functions (in the domain $\widetilde{J}_x^2>\widetilde{J}_z^2>\widetilde{J}_y^2$)
\begin{eqnarray}
\label{cb}
\widetilde{J}_x&=&\mathrm{ns}(\lambda,k)\nonumber\\
\widetilde{J}_y&=&\mathrm{cs}(\lambda,k) \\
\widetilde{J}_z&=&\mathrm{ds}(\lambda,k).\nonumber
\end{eqnarray}
Action on this solution by the group of transformations $\lambda\to\lambda+\psi$ and  $\lambda\to A\lambda,\widetilde{J}\to \widetilde{J}/A$ gives the 3-parameter family of solutions \cite{shiba,yosida}
\begin{eqnarray}
\label{c6}
\widetilde{J}_x&=&A\cdot\mathrm{ns}(A\lambda+\psi,k)\nonumber\\
\widetilde{J}_y&=&A\cdot\mathrm{cs}(A\lambda+\psi,k)\\
\widetilde{J}_z&=&A\cdot\mathrm{ds}(A\lambda+\psi,k), \nonumber
\end{eqnarray}
 wherefrom follows Eq. (\ref{amm}).
Permutation of the indices $x,y,z$ gives solutions in other domains.

\section{Let a hundred flowers blossom}

The two representations of the solution we got (through  Weierstrass and through Jacobi elliptic functions) in spite of looking differently, are equivalent.
Weierstrass elliptic function can be expressed through Jacobi one using equation
\begin{eqnarray}
{\cal P}(\lambda)=e_3+\frac{e_1-e_3}{\mathrm{sn}^2w},
\end{eqnarray}
where  $e_{1,2,3}$ are three roots of
the RHS of Eq. (\ref{we}), considered as a polynomial,
and where the modulus $k$ of the Jacobi function equals
\begin{eqnarray}
k\equiv \sqrt {\frac{e_2-e_3}{e_1-e_3}}
\end{eqnarray}
and it's argument $w$ equals
\begin{eqnarray}
w\equiv z\sqrt {e_1-e_3}.
\end{eqnarray}
Starting from Eq. (\ref{scalinga0g}) we get
\begin{eqnarray}
e_1=-RSy_0,\;\;e_2=-QSz_0,\;\;e_3=-QRx_0.
\end{eqnarray}
Thus for $Q=R=S=-1$ we recover Eq. (\ref{c6}).

\section{Special elliptic cones}
\label{elli}

To understand the  geometric meaning of the solution (\ref{amm}) let us start from  elementary geometry.
The Euclidean space ($x,y,z)$ can be in a unique way foliated into the elliptic cones of special ($\alpha+\beta+\gamma=0$) type
\begin{eqnarray}
\label{P}
\alpha x^2+\beta y^2+\gamma z^2=0.
\end{eqnarray}
(For $\alpha=0$ or $\beta=0$ the special cone is a pair of planes.)

The foliation includes three families of cones (the axis of the cones of each family is one of the Cartesian  axes).
These families will be referred to as $x$-cones, $y$-cones and $z$-cones.
An apex angle of a given cone is the angle between the cone's axis  and the  section of the cone  by a  plane
which contains the cone axis.
For example, for a $z$-cone, $\theta_{zx}$ is
the angle  between the OZ axis  and the  section of the cone  by the plane $x=0$. It is obvious that
\begin{eqnarray}
\label{P2}
\cos\theta_{zx}=-\frac{\beta}{\gamma}.
\end{eqnarray}

Now let us go from geometry to calculus.  Recalling the identity, elliptic functions satisfy
\begin{eqnarray}
1-k^2+k^2\mathrm{cn}^2(\lambda,k)-\mathrm{dn}^2(\lambda,k)=0,
\end{eqnarray}
one realizes that the solution (\ref{amm}) satisfies
\begin{eqnarray}
(1-k^2)J_x^2+k^2J_y-J_z^2=0.
\end{eqnarray}
Hence the special cones in the phase space $J_x,J_y,J_z$  remain invariant under the evolution (this is why they were introduced above), and
the parameter  $k^2$ in the solution (\ref{amm}) is the cosine of the $\theta_{zx}$ apex angle of the special cone, the solution belongs to.
Note that the stable fixed point  $(J_x,J_y,J_z)=(0,0,0)$ of Eq. (\ref{scalinga00}) is
the  apex of all special cones \cite{brodsky}.

\section{Mathematical conclusions}

In this paper we have integrated  specific systems of $m$ coupled scaling equations in terms of known functions. Two specific features of the systems allowed us to do it.
First, it was possible to reduce each system to the form
\begin{eqnarray}
\label{mc}
\frac{d \widetilde{J}_n}{d\lambda}&=&R_n\frac{\Phi(\widetilde{J})}{\widetilde{J}_n},
\end{eqnarray}
where $\Phi$ is some function of the coupling constants, and $R_n$ are  constants. This allowed us to   obtain $m-1$ first integrals of the system. Geometrically this means that in the space with the coordinates $(x_1,\dots x_m)=(\widetilde{J}_1^2\dots \widetilde{J}_m^2)$ a flow line of Eq. (\ref{mc}) lies on an arbitrary ray in the direction $(R_1,\dots R_m$).

The integrals being found, we are left with a single differential equation  for a single variable (whatever function of the coupling constants is chosen as such variable). This equation is of the first order,  solved with respect to the derivative of the dependent variable, and has the RHS which does not contain the independent variable. It means that  a system which can be reduced to Eq. (\ref{mc}) (for any $\Phi$,  any $m$ and  any set of constants
$R_n$) can be integrated in quadratures.

Second, the function $ \Phi$ being what it was, we were able to write down the solutions in terms of known transcendental functions
(circular trigonometric and hyperbolic  for $m=2$ and elliptic  for $m=3$).
 Jacobi elliptic functions  appear when we chose a coupling constant  (or inverse coupling constant, or the ratio of two coupling constants, as we understand looking at Eq. (\ref{jac})) as the above mentioned variable. Weierstrass elliptic functions appear when  the square of a coupling constant is chosen.

\end{appendix}


\begin{thebibliography}{99}

\bibitem{anderson} P. W. Anderson, J. Phys. C {\bf 3}, 2439 (1970).

\bibitem{hewson} A. C. Hewson, {\it The Kondo Problem to Heavy Fermions}, (Cambridge University Press, Cambridge, 1993).

\bibitem{sengupta} K. Sengupta and G. Baskaran, Phys. Rev. B {\bf 77}, 045417 (2008).

\bibitem{wehling}   T. O.Wehling, A. V. Balatsky, M. I. Katsnelson, A. I. Lichtenstein,
and A. Rosch, Phys. Rev. B {\bf 81}, 115427 (2010).

\bibitem{vojta} M. Vojta, L. Fritz, and R. Bulla, Europhys. Lett. {\bf 90}, 27006 (2010).

\bibitem{uchoa} B. Uchoa, T. G. Rappoport, and A. H. Castro Neto, Phys. Rev. Lett.
{\bf 106}, 016801 (2011); {\bf 106}, 159901(E) (2011).


\bibitem{fritz} L. Fritz and M. Vojta,  Rep. Prog. Phys. {\bf 76}, 032501 (2013).

\bibitem{chen} J.-H. Chen, L. Li, W. G. Cullen, E. D.
Williams, and M. S. Fuhrer,  Nat Phys {\bf 7}, 535
 (2011).

\bibitem{vojta3} M. Vojta and R. Bulla, Eur. Phys. J. B {\bf 28}, 283 (2002).

\bibitem{zhuravlev} A. K. Zhuravlev, V. Yu. Irkhin,  Phys. Rev. B {\bf 84}, 245111 (2011)

\bibitem{kanao} T. Kanao, H. Matsuura, and M. Ogata, J. Phys. Soc.
Jpn. {\bf 81}, 063709 (2012).

\bibitem{cazalilla}  M. A. Cazalilla, A. Iucci, F. Guinea, and A. H. Castro
Neto,  arXiv: cond-mat/1207.3135.

\bibitem{mitchell2} A. K. Mitchell and L. Fritz,  Phys. Rev. B {\bf 88}, 075104 (2013).

\bibitem{mitchell} A. K. Mitchell, M. Vojta, R. Bulla, and L. Fritz, \prb {\bf 88}, 195119 (2013).

\bibitem{shirakawa} T. Shirakawa and S. Yunoki,  Phys. Rev. B {\bf 90}, 195109
(2014); ibid  {\bf 93}, 205124 (2016).


\bibitem{fradkin} D. Withoff and E. Fradkin, \prl {\bf 64}, 1835 (1990).

\bibitem{chen2} K. Chen and C. Jayaprakash, J. Phys.: Condens. Matter {\bf 7}, L491 (1995).

\bibitem{buxton} C. Gonzalez-Buxton and K. Ingersent, Phys. Rev. B {\bf 57}, 14254 (1998).

\bibitem{bulla} R. Bulla, Th. Pruschke, and A. C. Hewson,  J. Phys.: Condens.
Matter {\bf 9}, 10463 (1997).

\bibitem{bulla2}  R. Bulla, M. T. Glossop, D. E. Logan, and Th. Pruschke,  Journal of Physics: Condensed Matter {\bf 12}
4899 (2000).

\bibitem{glossop} M. T. Glossop and D. E. Logan, Eur. Phys. J. B {\bf 13}, 513 (2000).

\bibitem{fritz2} L. Fritz and Matthias Vojta, Phys. Rev. B {\bf 70}, 214427 (2004).


\bibitem{cox} D.L. Cox and A. Zawadowski, Adv. Phys. {\bf 47},  599 (1998).

\bibitem{irkhin} V. Yu. Irkhin, M.I. Katsnelson and A.V. Trefilov, Zh. Eksp.
Teor. Fiz. {\bf 105},  1733 (1994); Physics Letters A {\bf 213},  65 (1996).

\bibitem{nozieres} Ph. Nozieres and A. Blandin, J. de Phys. {\bf 41}, 193 (1980).

\bibitem{landau3} L. D. Landau and E. M. Lifshitz, {\it Landau and Lifshitz Course of Theoretical Physics: Vol. 3 Quantum Mechanics}, (Pergamon Press, 1991).

\bibitem{landau}  L. D. Landau, A. A. Abrikosov, and I. M. Khalatnikov, Dokl. Akad. Nauk SSSR {\bf 95}, 497  (1954).

\bibitem{pokrovskii} A. Z. Patashinskii and V. L. Pokrovskii, {\it Fluctuation Theory of Phase transitions, 2 ed.},  (Moskow. Nauka, 1982) [in Russian];
English translation of the 1 ed see: {\it Fluctuation Theory of Phase transitions},  (Pergamon Press, 1979)

\bibitem{ingersent} M. Cheng, T. Chowdhury, A. Mohammed, and K. Ingersent, preprint
arXiv:1702.06515.

\bibitem{landau2} L. D. Landau and E. M. Lifshitz, {\it Landau and Lifshitz Course of Theoretical Physics: Vol. 1  Mechanics}, (Elsevier, 1976).

\bibitem{shiba} H. Shiba, Prog.  Theor. Phys. {\bf 43}, 601 (1970).

\bibitem{abram} M. Abramowitz, I. A. Stegun eds., {\it Handbook of Mathematical Functions with Formulas, Graphs, and Mathematical Tables},
(National Bureau of Standards, Washington, 1964).

\bibitem{yosida} K. Yosida, {\it Theory of Magnetism} (Springer, Berlin Heidelberg New York, 1996).

\bibitem{brodsky}
I. A. Brodsky,
\begin{quote}
And Euclid did not know that  approaching the apex of the cone,\\
A thing is going to meet not it's end, but eternity.\\
\end{quote}
\begin{flushright}
-I sit by the window (1964)
\end{flushright}

\end{thebibliography}
\end{document}